10.2  COUPLED ATMOSPHERE-FIRE SIMULATIONS OF FIREFLUX:
IMPACTS OF MODEL RESOLUTION ON MODEL PERFORMANCE

A. K. Kochanski[1], S. K. Krueger[1], M. A. Jenkins[2], J. Mandel[3] and J. D. Beezley[3]
[1]Dept of Atmospheric Sciences University of Utah, Salt Lake City, UT,
[2]Depatment of Earth and Space Science and Engineering, Faculty of Pure and Applied Science York University, Toronto, Ontario, Canada,
[3]Department of Mathematical and Statistical Sciences, University of Colorado, Denver

## 1. Introduction

Even though the scales and intensities of grass fires are not as great as forest fires, grass fires can present serious threats to firefighters and communities. Due to their very high spread rates they may be difficult to confine, and when they run out of control they can severely affect communities located within grassland environments. The ability to forecast grass fire spread could be of a great importance for agencies making decisions about prescribed burns. The regular spot forecasts may not provide enough information to predict spatial fire behavior, especially in terrain where meteorological conditions are not uniform. Coupled atmosphere-fire models are good candidates for providing realistic fire spread forecasts (Clark et al., 1996; McGrattan et al., 2004; Linn, 2005; Coen, 2005), but their usefulness is limited by the time required for completing the coupled atmosphere-fire simulations. In this study we analyze the sensitivity of a coupled model with respect to the vertical resolution of the atmospheric grid and the resolution of fire mesh that both affect computational performance of the model. Based on the observations of the plume properties recorded during the FireFlux experiment (Clements et al., 2007), we try to establish the optimal model configuration that provides realistic results for the least computational expense.

## 2. Model description

In this study we use the WRF-Sfire which is a coupled atmosphere-fire model based on the Weather Research and Forecasting model (WRF). It combines the WRF dynamical core (ARW) with the fire module, which bases on the semi-empirical Rothermel model (Rothermel, 1972), and traces the fire line by the level set method (Mandel et al 2009 and 2011). The model is embedded into the WRF modeling framework allowing for an easy set up for idealized and real cases requiring realistic meteorological forcing and detailed description of the fuel types and topography. The nesting capabilities of WRF allow for running the model in multi-scale configurations, where the outer domain, run at relatively low resolution, resolves the large-scale synoptic flow, while the gradually increasing resolution of the inner domains allow for realistic representation of smaller scales, required for realistic rendering of the fire behavior. The WRF-Sfire utilizes two-way coupling between the atmosphere and the fire. The wind at the fire line simulated by WRF is used for computation of the rate of the fire spread, which allows for estimation of the amount of burnt fuel and the fire front propagation. The heat and moisture released at the surface by a fire are fed into WRF. As a result the atmosphere "feels the fire" and responds to it, by changing air temperature and humidity (see Fig. 1)

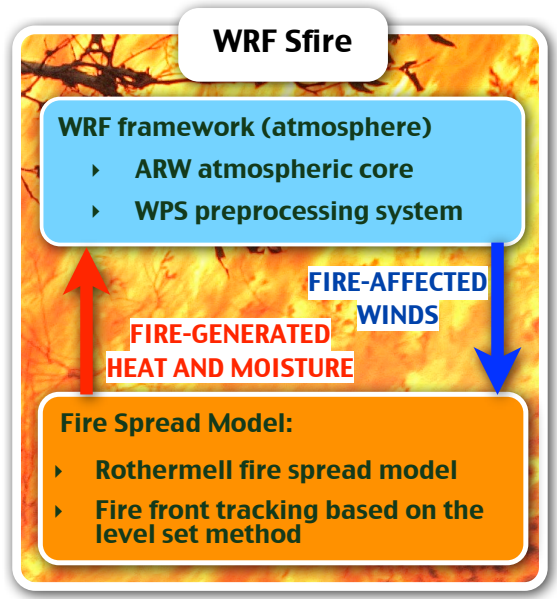

Fig. 1. Diagram of two-way coupling in the WRF-Sfire model.

Corresponding author address: Adam Kochanski,
Atmospheric Sciences Dept. of The University of Utah,
135 S 1460 E, 84112-0110, Salt Lake City, UT
E-mail: adam.kochanski@utah.edu

These changes in the temperature and moisture fields comprise dynamical forcing, that in turn affects the wind in the vicinity of the fire. This mechanism allows for driving the fire combustion and convection column with more realistic, fire-affected winds as opposed to the ambient winds, not affected by the presence of the fire, which are used one-way coupled models like FARSITE (Finney, 1998) or BEHAVE (Burgan & Rothermel, 1984). In order to keep the computational cost down, yet provide realistic representation of the fire front, and accurate estimation of the burnt area and fire-induced heat and moisture fluxes, the WRF-Sfire uses two grids. The coarser atmospheric mesh is used for the 3D computations of the atmospheric state, since they comprise the main computational cost associated with the fire spread simulation in the coupled atmosphere-fire framework. The less intensive 2D computations of the level set function (representing the fire front), as well as fluxes emitted by the fire, are performed on the finer fire mesh. The diagram showing two grids utilized by the WRF-Sfire is presented in Fig. 2.

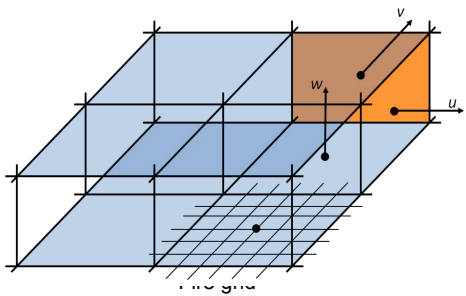

Fig. 2. Numerical grids of WRF-Sfire. The coarse atmospheric grid (3D) and the fine fire grid (2D) at the surface. Refinement ratio equal to 6 shown.

More details about the WRF-Sfire model used in this study can be found in (Mandel et al., 2011) and in previous work (Mandel et al., 2009).

**3. Experimental Setup**

To examine the impact of the changes in the model setup on the quality of the obtained results we simulate the passage of a grass fire as recorded during the FireFlux experiment (Clements et al., 2007). The diagram showing the FireFlux setup is presented in Fig. 3. The fire is ignited along the white dashed line, and the north-easterly wind advances the fire line south west. As the fire crosses the measurement towers MT and ST, temperatures, wind speed, and humidity are recorded.

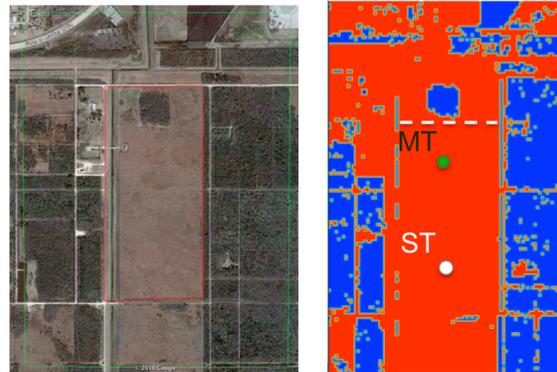

Fig. 3. Diagram of the FireFlux experiemnt:al setup. Aerial photo - left panel, Model domain with land use categories (red grass, blue forest), position of main and short measurment towers (MT, ST) and the ignition line (white dashed line) – right panel.

The model domain is presented in Fig. 3. The atmospheric grid has horizontal resolution of 10m and has 100x160 points. The model top is at 1200m above the ground. In each case we use 80 vertical levels, but we vary the height of the first level from 1m to 20m above the ground. These changes in the depth of the first model level are associated with changes in the number of the model levels within first 50m AGL, which vary from 2 to 22 (see Fig. 4).

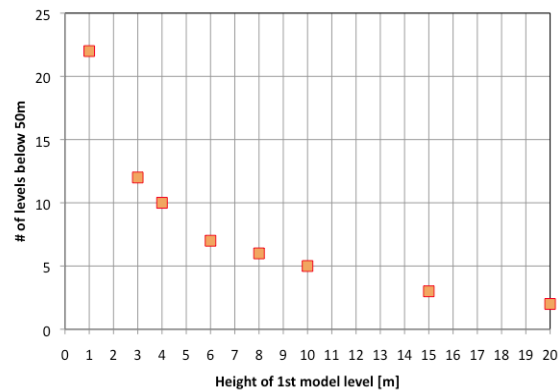

Fig. 4. Tested vertical model configurations - number of model levels below 50m as a function of the height of the 1st model level.

We test 6 different refinement ratios between the atmospheric and the fire meshes: 6, 10, 15, 20, 25 and 30, that translate into the fire mesh resolutions varying from 1.66m to 0.33m.

## 4. Results

*4.1 Impact of the fire model resolution on the computational cost of running the coupled model.*

The refinement of the fire grid is associated with an increase in the computational cost, corresponding to computations related to the fire propagation performed on an increasingly larger number of points. In order to assess the impact of the refinement ratio between the atmospheric and the fire mesh on the computational cost of the coupled atmosphere-fire simulation, we performed 6 simulations with identical set up of the atmospheric grid, but with different atmosphere-fire refinement ratios varying from 6 to 30. As can be seen in Fig. 5, the computational cost associated with running the fire model is strongly dependent on the refinement ratio.

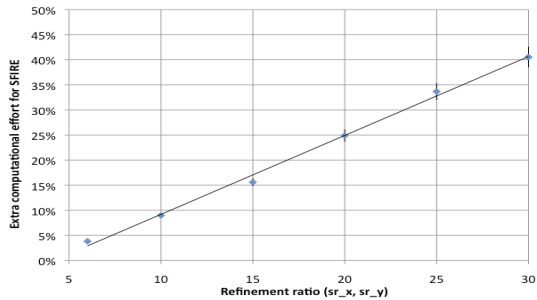

Fig. 5. Fire model resolution versus the computational cost.

For low fire-model resolutions the computational effort associated with the fire model is very modest. For refinement ratios below 10, the fire model accounts for less than 10% of the total computational cost. However, as the refinement ratio increases so does the computational cost of running the fire model. As a consequence, for the highest tested refinement ratio of 30 (fire model resolution of 0.33m), the fire model accounts for approximately 40% of the total computational cost. That means, that theoretically just by changing the refinement ratio between the fire and atmospheric grids in a 6-30 range, the computational cost and the model execution time may be reduced by over 35%. Of course decreasing the resolution of the fire model may also decrease the quality of the simulation and lead to unrealistic results. Therefore, for each tested refinement ratio we compare the computational results with FireFlux observations (see next section).

*4.2. Effects of the fire model resolution on the quality of the obtained results.*

As the refinement ratio increases the fire model should provide more detailed representation of the fire line shape as well as more accurate estimates of the burnt area, released heat and moisture. However, as discussed in the previous section, a refinement of the fire grid is associated with a serious increase in the computational cost. Therefore in this section we analyze the results from the simulations performed with different fire model resolutions in order to assess if and at what refinement ratio results start to converge, and what refinement ratio may be considered as optimal. We evaluate the results based on the wind speed and air temperature recoded by the main tower at 2m above the ground.

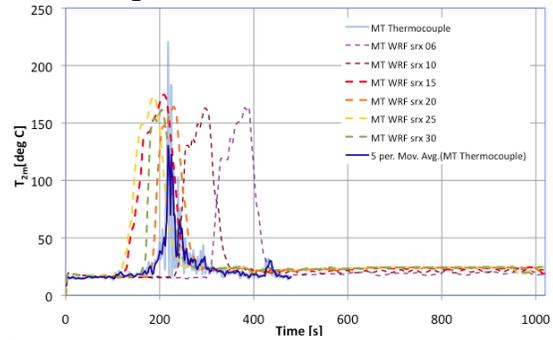

Fig. 6. Air temperate 2m above the ground recorded at the main tower (solid lines) and simulated (dashed lines). Light blue line - 1Hz observations, navy - measurements adjusted to the model output frequency (0.2Hz).

As can be seen in Fig. 6 changes in the refinement ratio affect both the timing of temperature peak (corresponding to the arrival of the fire front), as well as its magnitude. The simulations performed with the refinement ratio of 10 and 6 significantly departure from the observations and other simulations performed with finer fire meshes. The simulations with coarse fire grid lose their realism due to significant underestimation of the fire rates of spread that lead to the delay in the simulated fire front arrival by up to 180s. A possible reason for the observed relationship between the refinement ratio and the simulated rate of spread is the accuracy of the computation of the burnt area and heat fluxes. Since the fire front has an approximately parabolic shape, its coarse multi-linear representation must lead to an underestimation of the burnt area. This in turn means under prediction of the released heat, and the plume rise. The underestimation in the plume vertical velocity is associated with reduced convergence at the surface and lower magnitudes of the fire induced winds at the surface. The analysis of the wind speed at 2m above the ground presented in Fig. 7, shows the impact of fire model resolution on the magnitude of the wind speed at 10m above the fire front.

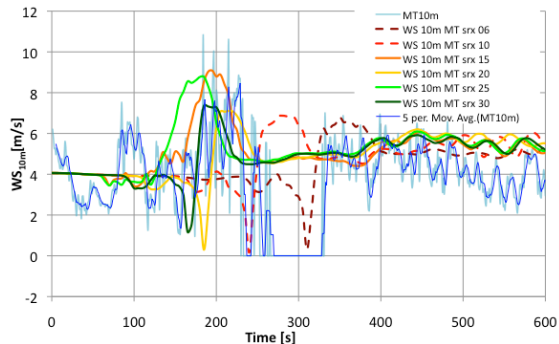

Fig. 7. 10m wind speed at the main tower. Solid blue line - measurement data at 1Hz, solid navy line - measurements adjusted to model output frequency of 0.2Hz. Other colors show model results from runs with various refinement ratios.

It is evident that lower fire model resolutions (lower refinement ratios) are associated with lower than observed peak wind speeds. This confirms that the underestimated burning area, due to too coarse a representation of the fire front, may be responsible for the underestimation of the fire-emitted heat flux, fire plume velocities and consequently fire-induced surface wind. Since the fire model is driven by the local wind speed at the fire line, its underestimation may lead to an unrealistically low fire rate of spread observed in Fig. 6 and Fig. 7. Looking at both these graphs, it is evident that results for refinement ratios above 10 do not differ substantially. For refinement ratios of 15 and more all results start to merge. It seems that even for relatively coarse refinement ratio of 15, obtained results a relatively good, and using a finer fire mesh does not improve results substantially.

### 4.3 Impact of the vertical setup of the atmospheric model on the computational cost

The horizontal resolution of the model is not the only factor affecting the model computational performance. In fact, since the vertical velocities within fire plumes are substantial, in fire simulations they often become limiting factors in terms of maximum numerically stable time steps. Measurement data collected during the FireFlux experiment (Clements et al., 2008) showed that maximum updraft velocities within the grass fire plume may reach 8 m/s within the first 50m above the ground. An adequate vertical resolution within this layer is important in terms of capturing the atmosphere-fire interaction, and rendering the fire-induced wind acceleration at the surface. Therefore, the vertical model setup near the surface may be a key factor affecting both the computational performance (trough the maximum numerically stable time step) and the quality of obtained results (through the degree to which the fire-atmosphere coupling is captured). The fire plume updraft velocity changes with height, as does the depth of the model layers in the stretched grid setup. In order to evaluate which model levels are most prone to violating the vertical CFL (Holton 1992) criteria, for each vertical model setup we computed the maximum allowable time step. For each vertical model setup (see Fig. 4) we compute the maximum vertical velocity at each model level and divide it by the depth of model layer at that level. Dividing the depth of the model layer by the maximum updraft observed in that layer gives the maximum numerically stable time step corresponding to the vertical CFL=1 for that model layer. The maximum stable time steps computed for all model levels at various vertical configurations are presented in Fig. 8.

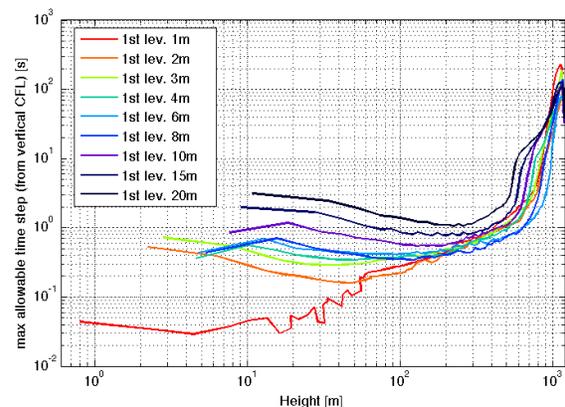

Fig. 8. Maximum numerically stable time step computed for each model level, for various vertical model configurations.

The vertical model setup limits therefore the maximum numerically stable time step. The configurations with the first model levels closer to the surface and higher density of the model levels at the bottom require significantly lower time steps to handle strong fire-induced updrafts. That translates into higher computational cost and significantly longer execution times. As the density of model levels at the surface decreases, the height that limits the maximum numerically stable time step (where the minimum of the stable time step occurs) increases, and the model theoretically may be run faster (with a longer time step). Even though in every case the height limiting the time step is below 100m AGL (above ground level), the minima in the allowable time steps occur at various heights. For the configurations with the first model layer placed 1m AGL, the time step is limited by the updraft in the layer around 5m AGL. As the density of the model levels in the lower part of the domain decreases, the height at which the time step minimum occurs increases, and for the setup with the first model level at 20m AGL, the minimum in allowable time step is associated with relatively high layers located between 110 and 120m AGL. Fig. 8 also suggests that the vertical configurations with the first model layers at 1 and 2m AGL are not practical performance-wise since they require extremely short time steps. By moving the first model level from 1 to 3m AGL the time step may be increased by a factor of 10. A further decrease in the model vertical resolution at the surface does not provide such a significant computational gain. The configurations with the first model level between 4 and 8 AGL require similar time steps (around 0.3-0.5s) and in order to further increase the time step, the first model level would have to be moved up above 15m AGL. As discussed in this section, the vertical setup of the model has a big impact on the maximum numerically stable time step and the model performance, and clearly decreasing the model resolution at the surface reduces computational cost by allowing for longer time steps. However, too coarse a vertical resolution may impede the model's ability to capture the atmosphere-fire interaction and lead to unrealistic results. Therefore in the next section we analyze the impact of the vertical model setup on the model results.

### 4.4. Effects of the vertical model setup on the quality of the obtained results.

In order to evaluate the impact of the changes in the vertical model resolution on the model results, we compare the simulated wind speed and temperature to the data collected at the main tower during the FireFlux experiment. The analysis of the simulated wind speed at 10m AGL suggests that only the vertical setups with the first model level placed below 6m above the surface are able represent realistically the fire-induced wind speed up. As shown in

Fig. 9, the vertical configurations with the first model levels placed between 1 and 6m above the ground show practically the same timing of the wind speed peak.

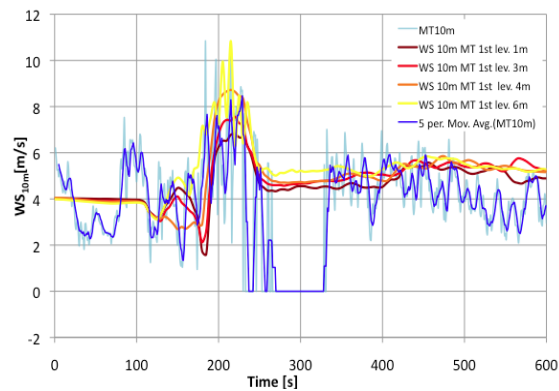

Fig. 9. Time series of the simulated and observed wind speed at 10m AGL. The light blue line shows 1Hz measurement data, the navy line shows measured wind speed adjusted to the model output frequency of 0.2Hz. Other lines present simulated wind speed for various vertical model configurations with the first model placed from 1m to 6m above the ground.

That suggests that in all these cases the fire rate of spread is very similar. However, the magnitudes of the wind speed maxima differ, and as the depth of the first model layer increases so does the 10m wind speed. This is not a rule though. As shown in Fig. 10, for coarser vertical model resolutions at the surface (first model level above 8m AGL), there is no clear relationship between the vertical resolution and simulated wind speed. In fact, for coarser vertical resolutions the results show an opposite trend. The peak wind speed for the configuration with the first model level at 8m AGL is significantly greater than for any other setup. Unlike in the finer resolution cases, results from simulations performed with lower vertical resolutions show that the timing of the fire wind speed maximum is also dependent on the vertical model setup.

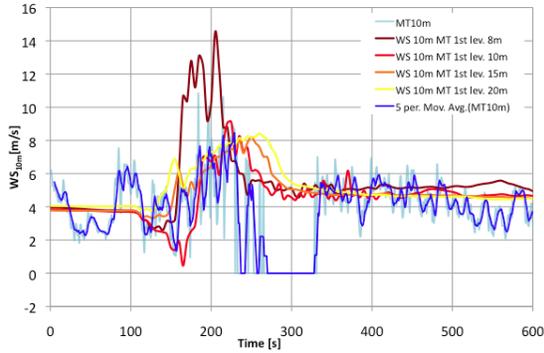

Fig. 10. Time series of the simulated and observed wind speed at 10m AGL. The light blue line shows 1Hz measurement data, the navy line shows measured wind speed adjusted to the model output frequency of 0.2Hz. Other lines present simulated wind speed for various vertical model configurations with the first model placed from 8m to 20m above the ground.

The temperatures simulated by the model with various vertical resolutions do not differ as much as the wind speed. All the simulations with the first model level at 6m and below show very similar results both in terms of the magnitude of the temperature rise associated with the fire front passage and its timing (see Fig. 11a). The amplitudes of the temperature peaks are practically identical and the timings of the plume passage by the measurement tower are captured very well. However, for coarser vertical model grids the differences are more pronounced. As shown in Fig. 10 b, for the cases run with coarser vertical grids (first model levels 8m AGL and above) both the magnitude and the timing of the temperature peaks differ. For the cases with the depths of the first model layer of 15 and 20m, the maximum simulated temperatures reach only 50% of the observed values, and the peaks appear too late compared to observations. For these cases the 10m values of the wind speed and air temperature are reconstructed based on the Monin-Obukhov similarity theory and not computed directly by the model, since its first level is more than 10m AGL. That may partially explain the underestimation in the simulated temperature maxima evident in Fig. 11b. Another factor that may impact the magnitude of the temperature peak as well as it's timing, is the model's ability to capture the coupling between the fire and the atmosphere and the resulting surface wind acceleration. If the heat emitted by the fire is distributed over a deep first model layer, the increase in the temperature within this layer will be lower than for a case when the same heat is distributed over a shallow layer.

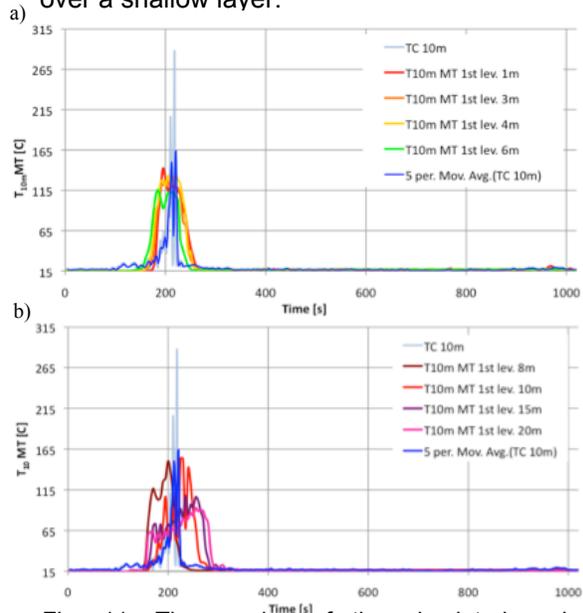

Fig. 11. Time series of the simulated and observed air temperature at 10m AGL. The light blue line shows 1Hz measurement data, the navy line shows measured air temperature adjusted to the model output frequency of 0.2Hz. Other lines present simulated air temperature for various vertical model configurations with the first model placed from 1m to 20m above the ground.

## 5. Summary

Both, the resolution of the fire model (defined by the refinement ratio), as well as the vertical setup of the atmospheric component, strongly affect the computational performance of the coupled model. The refinement of the fire model mesh is associated with an increase in the number of points for which the fire computations are performed and consequently also an increase in the computational cost. An increase in the vertical model resolution, even at the same total number of vertical atmospheric levels, also makes simulations more computationally intensive. This is due to the fact that in order to resolve strong updrafts associated with fire plumes on a finer vertical mesh, the time step must be significantly reduced.

Our test simulations showed that the contribution of the fire model to the total cost of running the coupled model rises quickly from a modest 4% for the refinement ratio of 6 up to 40% for the fine fire grids refined by a factor of 30 with respect to the atmospheric grid. Changing the depth of the first model layer from 1m to 3m allows for a huge increase in the time step and a reduction of the computational cost by a factor of 10.

In fire spread forecasting cases, requiring high computational performance, the model setup must be designed very carefully. From the point of view of the computational cost, a relatively coarse fire mesh and a low vertical resolution of the atmospheric model are desirable. On the one hand, lowering the resolution of the fire and atmospheric vertical grid leads to significant deterioration of the quality of the results. An analysis of the simulated wind speed and temperature associated with the fire front passage suggests that both too coarse fire grid resolution as well as too coarse vertical resolution may impede the results. The errors in the simulated rate of spread for various fire grid resolutions presented in Fig. 12 show that in order to capture the fire propagation characteristics correctly, the refinement ratio must greater than 15. Nevertheless, a further increase in the refinement ratio above 25 does not seem practical, since it increases the computational cost significantly but does not improve the results. The model results with refinement ratios above 15 converge nicely, suggesting that the refinement ratio of 20 should be probably considered as optimal, since it provides best results, yet at an acceptable cost.

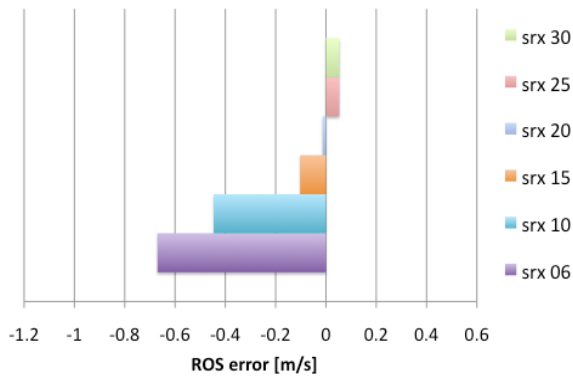

Fig. 12. Error in fire rate of spread prediction for various refinement ratios between the atmospheric and fire grids.

A similar analysis of the error of the rate of spread prediction for various vertical configurations also allow provide information on optimal configurations. Results presented in Fig. 13 show that increasing the vertical resolution and lowering the first model level generally improves the results, however for a very shallow first model level (1m) the results are actually not much better than for a case with the first model level at 8m AGL. This setup overestimates the wind speed at the very surface but also underestimates it at 5m AGL. Since the wind speed used by the fire model is taken at 6.1m AGL, this setup underestimates the rate of the fire spread. Nevertheless, all other configurations with the 1st model levels located between 2 and 6m above the ground provided very similar results in terms of the rate of spread. The set up with a 6m-depth initial model layer underestimated the peak temperature at 10m AGL (see Fig. 11 a) so probably the slightly refined setup with the 1st model level at 4m AGL may be considered as optimal.

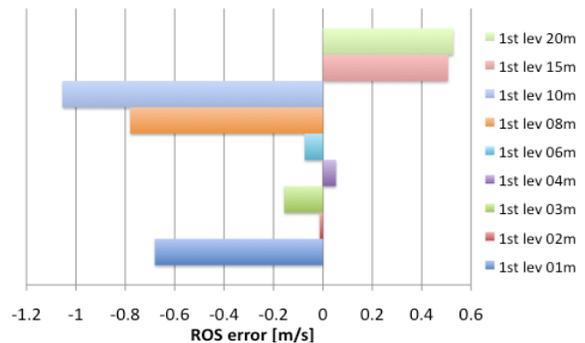

Fig. 13. Error in the simulated fire rate of spread for various configurations of the vertical model grid.

The sensitivity study presented in this work suggests that the setup of the coupled atmosphere-fire model has a big impact on the model computational performance as well as the quality of obtained results. However, the optimal configurations pointed out here shouldn't be treated as universal. Different horizontal resolutions of the atmospheric model, or complex fuels may require finer fire grids and higher vertical resolutions. More numerical experiments for more realistic fires are required to better understand the impact of the model configuration on its accuracy.